\documentclass[12pt,preprint]{aastex}

\slugcomment{}

\shorttitle{Lensed QSO SDSS J0924+0219}
\shortauthors{Inada et al.}

\begin{document}

\title{SDSS~J092455.87+021924.9: an Interesting 
Gravitationally Lensed Quasar from the Sloan Digital Sky Survey}

\author{
Naohisa Inada,\altaffilmark{1}
Robert H. Becker,\altaffilmark{2,3}
Scott Burles,\altaffilmark{4}
Francisco J. Castander,\altaffilmark{5,6}
Daniel Eisenstein,\altaffilmark{7}
Patrick B. Hall,\altaffilmark{8,9}
David E. Johnston,\altaffilmark{10}
Bartosz Pindor,\altaffilmark{8}
Gordon T. Richards,\altaffilmark{11}
Paul L. Schechter,\altaffilmark{4,12}
Maki Sekiguchi,\altaffilmark{1}
Richard L. White,\altaffilmark{13}
J. Brinkmann,\altaffilmark{14}
Joshua A. Frieman,\altaffilmark{10,15}
S.J. Kleinman,\altaffilmark{14}
Jurek Krzesi\'{n}ski,\altaffilmark{14,16}
Daniel C. Long,\altaffilmark{14}
Eric H. Neilsen, Jr.,\altaffilmark{15}
Peter R. Newman,\altaffilmark{14}
Atsuko Nitta,\altaffilmark{14}
Donald P. Schneider,\altaffilmark{11}
S. Snedden,\altaffilmark{14}
and Donald G. York\altaffilmark{10,17}
}

\altaffiltext{1}{Institute for Cosmic Ray Research, University of Tokyo, 5-1-5 Kashiwa, Kashiwa City, Chiba 277-8582, Japan.}
\altaffiltext{2}{Physics Department, University of California, Davis, CA 95616.}
\altaffiltext{3}{IGPP-LLNL, L-413, 7000 East Avenue, Livermore, CA 94550.}
\altaffiltext{4}{Physics Department, Massachusetts Institute of Technology, 77 Massachusetts Avenue, Cambridge, MA 02139.}
\altaffiltext{5}{Department of Physics, Yale University, PO Box 208121, New Haven, CT 06520.}
\altaffiltext{6}{Departamento de Astronom\'{\i}a, Universidad de Chile, Casilla 36-D, Santiago, Chile.}
\altaffiltext{7}{Steward Observatory, University of Arizona, 933 North Cherry Avenue, Tucson, AZ 85721.}
\altaffiltext{8}{Princeton University Observatory, Peyton Hall, Princeton, NJ 08544.}
\altaffiltext{9}{Departamento de Astronom\'{\i}a y Astrof\'{\i}sica, Pontificia Universidad Cat\'{o}lica de Chile, Casilla 306, Santiago 22, Chile.}
\altaffiltext{10}{Department of Astronomy and Astrophysics, The University of Chicago, 5640 South Ellis Avenue, Chicago, IL 60637.}
\altaffiltext{11}{Department of Astronomy and Astrophysics, The Pennsylvania State University, 525 Davey Laboratory, University Park, PA 16802.}
\altaffiltext{12}{Institute for Advanced Study, Einstein Drive, Princeton, NJ 08540.}
\altaffiltext{13}{Space Telescope Science Institute, 3700 San Martin Drive, Baltimore, MD 21218.}
\altaffiltext{14}{Apache Point Observatory, P.O. Box 59, Sunspot, NM 88349.}
\altaffiltext{15}{Fermi National Accelerator Laboratory, P.O. Box 500, Batavia, IL 60510.}
\altaffiltext{16}{Obserwatorium Astronomiczne na Suhorze, Akademia Pedagogicazna w Krakowie, ulica Podchor\c{a}\.{z}ych 2, PL-30-084 Krak\'{o}w, Poland.}
\altaffiltext{17}{Enrico Fermi Institute, The University of Chicago, 5640 South Ellis Avenue, Chicago, IL 60637.}

\begin{abstract}

We report the discovery of a new gravitationally lensed quasar from
the Sloan Digital Sky Survey, SDSS~J092455.87+021924.9 (SDSS~J0924+0219). 
This object was
selected from among known SDSS quasars by an algorithm that was
designed to select another known SDSS lensed quasar
(SDSS~1226$-$0006A,B). Five separate components, three of which are
unresolved, are identified in photometric follow-up observations
obtained with the Magellan Consortium's 6.5m Walter Baade telescope at
Las Campanas Observatory. Two of the unresolved components
(designated A and B) are confirmed to be quasars with $z=1.524$; the
velocity difference is less than 100 km sec$^{-1}$ according to
spectra taken with the W. M. Keck Observatory's Keck II telescope on
Mauna Kea. 
A third stellar component, designated C, has the colors of a quasar with
redshift similar to components A and B. The maximum separation of the
point sources is $1\farcs78$. The other two sources, designated G and
D, are resolved. Component G appears to be the best candidate for the
lensing galaxy. Although component D is near the expected position of
the fourth lensed component in a four image lens system, its properties are
not consistent with being the image of a quasar at $z\sim1.5$.
Nevertheless, the identical redshifts of components A and B and the
presence of component C strongly suggest that this object is a
gravitational lens. Our observations support the idea that a foreground object 
reddens the fourth lensed component and that another unmodeled effect
(such as micro- or milli-lensing) demagnificates it, but 
we cannot rule out the possibility that SDSS0924+0219 is an example of
the relatively rare class of ``three component'' lens systems.

\end{abstract}


\section{Introduction}

If the rate of gravitational lenses among known quasars is roughly
0.1\% of all quasars (Turner, Ostriker \& Gott 1984), the expected
number of new gravitationally lensed quasars in the
Sloan Digital Sky Survey's (SDSS; York et al. 2000) final
spectroscopic quasar sample will be approximately $10^{2}$.
Furthermore, the entire SDSS photometric sample may contain an order
of magnitude more lensed systems (since the photometric observations
reliably probe $\sim2$ mag fainter than the spectroscopic sample).

The SDSS is a project to conduct parallel photometric and
spectroscopic surveys of 10,000 deg$^{2}$ of the sky centered
approximately on the North Galactic Pole, using a dedicated wide-field
2.5-m telescope at Apache Point Observatory (APO) in New Mexico, USA.
Photometric observations are done in five broad optical bands ($u$,
$g$, $r$, $i$, and $z$, centered at 3561{\,\AA}, 4676{\,\AA},
6176{\,\AA}, 7494{\,\AA}, and 8873{\,\AA} respectively (Fukugita et
al. 1996; Stoughton et al. 2002). The imaging camera consists of a
5$\times$6 array of large photometric CCDs (2048$\times$2048 pixels)
and 24 astrometric CCDs (2048$\times$128 pixels) (Gunn et al. 1998).
The imaging data are reduced by the photometric pipeline (Lupton et
al. 2001) using information from the astrometric pipeline (Pier et
al. 2002) and the 0.5-m photometric calibration telescope (Hogg et
al. 2001; Smith et al. 2002). Spectroscopic observations are done
with a multi-fiber spectrograph covering 3800{\,\AA} to 9200{\,\AA}
with the resolution R$=$1800.

We are interested in using the SDSS to find new lensed quasars since they have
proven to be useful for cosmological tests, especially for the
measurement of the cosmological constant by the number counts of
lensed quasars (Fukugita \& Turner 1991) as well as the measurement of the
Hubble constant by observations of the time delay between multiple
components (Refsdal 1964). Finding new lensed quasars in a large, homogeneous
survey, such as the SDSS will contribute greatly to the statistics of
lensed quasars (useful for measurements of the cosmological constant) and the
determination of their time delays (useful for constraining the Hubble
constant). To date the SDSS has yielded one new two-image lensed quasar, SDSS
1226$-$0006A,B (Inada et al. 2002) and here we report on the second
lensed quasar system discovered among the SDSS quasars.

To maximize the likelihood of discovering additional lens systems, we
are studying the SDSS parameters of previously known lensed quasars.
In particular, we studied the parameters of SDSS J1226$-$0006, and
using these parameters we developed an algorithm to select SDSS
1226-like objects from the SDSS database. This algorithm should be
sensitive to lensed quasars with separations on the order of
$1\farcs0$ to $2\farcs5$. Applying this algorithm to the
approximately 10,000 SDSS quasars discovered prior to 2001 December 1,
we identified five lensed quasar candidates. By definition one of them is SDSS
1226$-$0006. Another candidate is SDSS~J0924+0219 (09$^{h}$ 24$^{m}$
55$_{^{\verb+.+}}^{s}$87, +02$^{\circ}$ 19$^{'}$ $24\farcs9$, J2000)
which is identified as a $z=1.524$ quasar in the SDSS database. We
obtained photometric follow-up observations of SDSS~J0924+0219 using
the Magellan Consortium's Walter Baade 6.5-m (WB6.5m) telescope at Las
Campanas Observatory under good seeing conditions
($0\farcs55{\sim}0\farcs75$). Additional spectroscopic observations
of the components that were resolved by the WB6.5m images were
obtained with the Keck II telescope at the W. M. Keck Observatory on
Mauna Kea in $<1\farcs0$ seeing. Observations of the three
other candidates from this sample have been taken with the WB6.5m
telescope and indicate that these systems are also likely to be
lensed quasars; these systems will be discussed in future papers.

Section 2 of this paper briefly describes the new algorithm which led
to the discovery of SDSS0924+0219. Section 3 describes the follow-up
observations and show the results of them.
In \S~4, we discuss some of the interesting aspects of this
system. Finally, we present a summary of this paper in \S~5.

\section{Selection Method of Finding lensed quasars from the SDSS Quasar Catalog}

We now describe the manner in which SDSS~J0924+0219 was selected as a
gravitational lens candidate based upon the object parameters in the
SDSS object catalog (Lupton et al. 2001). First, we selected objects
that were confirmed to be quasars in the SDSS spectroscopic survey
(see Richards et al. 2002 for details of the SDSS quasar target
selection algorithm). We rejected quasars whose redshifts are less
than 0.6 since many low-redshift quasars are extended objects
(Schneider et al. 2002), making it hard to distinguish them from
unresolved lensed quasars. Next, we restricted our lensed quasar candidate sample
using some SDSS catalogued parameters, specifically the galaxy profile
fitting likelihood\footnote{We mainly used two SDSS catalogued parameters, 
``dev\_L'' (de Vaucouleurs profile fitting likelihood) and ``exp\_L'' 
(exponential disk fitting likelihood).}.
As the SDSS data are passed through the data reduction pipelines, each
extended object is fitted with a set of possible galaxy profiles
(Lupton et al. 2001) and labeled with the likelihood that each profile
explains the data. These likelihoods are useful for searching for
extended quasars, so we optimized our search criteria to use these
values, based on our study of the first SDSS lensed quasar, SDSS J1226$-$0006.

This algorithm targets lensed quasars whose separations are approximately
$1\farcs0{\sim}2\farcs5$, because we empirically confirmed that lensed quasars
whose separations are less than 1\farcs0 do not have large
``extended'' parameters in the SDSS object catalog; we cannot
distinguish these small separation lensed quasars from single, unresolved
quasars. For lensed quasars that have separations of more than 2\farcs5, each
lensed component should appear as a separate entity in the SDSS
catalog. As a result of applying this algorithm to the approximately
10,000 SDSS quasars (in $\sim1,100$ deg$^2$), we selected five lensed quasar
candidates. One of these five candidates is the first SDSS lensed quasar, SDSS
1226$-$0006, and another of these candidates is SDSS~J0924+0219. If
the total lensing rate is roughly 0.1\% of all quasars, then on the
order of $10$ lensed quasars are expected from the approximately
$10,000$ SDSS quasars that were known at the start of this work.
About half of the lensed quasars should have $1\farcs0{\sim}2\farcs5$
separations (Chiba \& Yoshii 1999); therefore, the result of this
algorithm is consistent with the theoretical estimate. Whether or not
this selection algorithm is the optimal way to select moderate
separation lens candidates from the SDSS imaging data remains to be
seen, since it could be biased towards SDSS J1226$-$0006-like lens
systems. However, the successful discovery of two lensed quasars
suggests that it is a reasonable method to use for our initial lens
search.

\section{Data analysis}

\subsection{Follow-up Observations}

Photometric follow-up observations of SDSS~J0924+0219 were obtained
using the WB6.5m telescope. The data were taken on 2001 December 15
with $u$, $g$, $r$, and $i$ filters using the Magellan Instant Camera
(MagIC, a 2048$\times$2048 CCD camera); the seeing was
0\farcs55${\sim}$0\farcs75 FWHM. The pixel size was 0\farcs069. The
exposure time was 300 sec in each band. Each CCD frame was
bias-subtracted and flat-field corrected.

Additional spectroscopic data were taken on 2002 January 12 with the
Keck II Echellette Spectrograph and Imager (ESI; Sutin 1997), using
the MIT-LL 2048$\times$4096 CCD camera and a 175 line mm$^{-1}$
grating. We used the echellette mode. The resolution of the
echellette mode of this spectrograph is 11.4 km sec$^{-1}$
pixel$^{-1}$. The spatial resolution scale of this spectrograph is
0\farcs153 pixel$^{-1}$. The spectral range covers 3900{\,\AA} to
11,000{\,\AA}. The exposure time was 1200 sec. 
We set the slit direction so that two of the three
components (components A and B, see below) were on the slit at the
same time. 
The two components are separated by 1\farcs78 and the
seeing was less than 1\farcs0 FWHM; the two components are clearly
distinct in the 2-dimensional ESI image. 
The two spectra were extracted separately using the usual method of 
summing the flux in a window around each object and subtracting sky 
from neighboring windows on either side of the trace. 
The only difference from a simple single
spectrum extraction was that we were careful to exclude the other
object from the sky windows. There was no need (and no attempt) to fit
the two spectra simultaneously because we cannot see any overlap along the
slit in the 2-dimensional ESI image. However, flux of component D (see below) 
actually affect the component A spectrum (we set the slit width 1\farcs0). 
We could not see it in the 2-dimensional ESI image, because component D is close to the component A, and
is more than two magnitudes fainter than component A. The estimated contaminations are 6.0\% of the flux of 
component A in $g$ band (4000\AA$\sim$5500\AA), 6.8\% in $r$ band (5500\AA$\sim$7000\AA), 
and 7.4\% in $i$ band (7000\AA$\sim$8500\AA), respectively.

We used a single slit position that included components A and B, 
which puts component G (see below) slightly off to one side. Since it is
much fainter (about 2 mag fainter than component A), 
it is not noticeable in the spectrum.

\subsection{Photometry}

The SDSS images (sky and bias subtracted and flat-field corrected) are
shown in all bands in Figure~1. Total magnitudes of the five
components are 18.68$\pm$0.02, 18.43$\pm$0.01, 18.34$\pm$0.01,
18.09$\pm$0.01, and 17.98$\pm$0.03 in $u^*$, $g^*$, $r^*$, $i^*$, and
$z^*$, respectively\footnote{The filters of the SDSS 2.5-m
are referred to as $ugriz$, while the still preliminary
2.5-m filter-based photometry will be called
${u^*}{g^*}{r^*}{i^*}{z^*}$ (Stoughton et al. 2002)}. These errors in
the magnitudes are statistical errors.

We show the full follow-up MagIC $r$ image including nearby stars
(star a, star b, and star c) as well as SDSS~J0924+0219 in Figure~2.
The non-PSFs subtracted (hereafter ``original'') $u$, $g$, $r$, and $i$
images of SDSS~J0924+0219 and the PSF-subtracted images of each band
are shown in Figure~3. The upper panels are the original images
and the lower panels are the PSF-subtracted images.
We subtracted PSFs using stars from the original images in all bands.
We used star a for the $u$ image, star b for the $r$ and the $i$
images and star c for the $g$ image, respectively. The ``peak'' flux and
the center coordinates of stars a, b, and c and components A, B, and C
were calculated by the single Gaussian fit (these results agree with 
the results obtained by the ``imexamine'' task in IRAF\footnote{IRAF is
the Image Reduction and Analysis Facility, a general purpose software
system for the reduction and analysis of astronomical data. IRAF is
written and supported by the IRAF programming group at the National
Optical Astronomy Observatories (NOAO) in Tucson, Arizona. NOAO is
operated by the Association of Universities for Research in Astronomy
(AURA), Inc. under cooperative agreement with the National Science
Foundation.}). We named the three stellar components ``components
A--C'' according to their magnitudes, the center extended object
``component G'' and the unknown component which remains after
subtracting PSFs ``component D''. The flux ratios between components
A and B are 0.47, 0.44, 0.43, and 0.43 in $u$, $g$, $r$, and $i$,
respectively, and those between components A and C are 0.44, 0.43,
0.41, and 0.40. In the original $i$ image one can see components G and
D, both are much more prominent in the PSF-subtracted images (except
for $u$). We show the reduced ${\chi}^{2}$ from the PSF subtraction
in Table~1. The reduced ${\chi}^{2}$ of the $g$, $r$, and $i$ images
of component A and component C are large, because component A has
contamination from components G and D, and component C has
contamination from component G. We cannot see components D and G in
the $u$ PSF-subtracted image. This fact suggests that component D is
not a quasar but rather a galaxy. We give our estimated magnitudes
and colors of the three stellar components, component G and component
D in Table~2. We used stars a, b, and c as the photometric standard
stars using their SDSS catalog magnitudes and positions.

\subsection{Spectra}

Spectra of components A and B taken with ESI on Keck II are shown in
Figure~4. \ion{Al}{3}, \ion{Si}{3]}, \ion{C}{3]}, and \ion{Mg}{2}
emission lines are seen clearly; both components are quasars at the
same redshift ($z=1.524$). The velocity difference between the two
components is less than 100 km sec$^{-1}$, calculated using the
\ion{Mg}{2} emission lines. The widths of the emission lines of both
components are also in good agreement. We do not have a spectrum of
component C, but the colors of this component indicate that this
component is also a quasar. A photometric redshift can be computed
for component C following Richards et al. (2001), assuming an $i-z$
color of $-0.025$. The resulting photometric redshift is
1.33$\pm$0.20, which is consistent with the redshifts of
components A and C, $z=$1.524. The redshifts and the widths of the
emission lines are summarized in Table~3.

\subsection{Astrometry}

We calculated the celestial coordinates of components A, B, and C based
on the SDSS celestial coordinates of the three stars (star a, star b, 
and star c, Figure~2) common to the MagIC $r$ image and the SDSS data.
The separations between components A and B, B and C, and A and C are
calculated to be 1\farcs78$\pm$0\farcs04, 1\farcs50$\pm$0\farcs03, and
1\farcs14$\pm$0\farcs04, respectively. We also calculated the
celestial coordinates of component G and component D after subtracting
PSFs. The results are summarized in Table~4.

\section{Discussion}

SDSS~J0924+0219 is certainly a lensed quasar because of its
morphology (there is a galaxy between three almost same 
color quasars) 
and the small velocity difference and the small line width
differences between components A and B. However, even though there
are no emission line profile differences between components A and B,
the ratio of the spectra of the two quasars is not constant with
$\lambda$; it steadily decreases from 0.7 at 4000{\,\AA} to 0.2 at
9000{\,\AA} as shown in Figure~5. The flux of component D would affect
the spectrum of component A, but the estimated contaminations are not 
large (see \S~3.1) and it would not give a dramatically change in the 
spectrum of component A. The difference in the flux density
ratio between the two components indicates that the two quasars might
not be from the same physical source. It is possible that the slit
used for the Keck spectroscopic observation was not precisely aligned
along the line between components A and B; combined with differential
refraction, this could have produced a spectroscopic flux density
ratio different from the photometric flux ratio. However, this cannot
be the full explanation, since it cannot explain why the spectroscopic
flux ratio is not smooth (Figure~5).

Another possibility for the differences in the spectra of the two
components is that continuum variations of the source quasar combined
with the differential time delay causes the wavelength dependence in
the flux density ratio. Continuum variations are often seen in
quasars (Tr\'{e}vese, Kron, \& Bunone 2001), and differences in flux ratios
between lensed component continua are seen in some catalogued lensed quasar
systems, e.g. HE 1104$-$1805 (Wisotzki et al. 1995) and FBQ 0951+2635
(Schechter et al. 1998)\footnote{Differential extinctions might also
cause the differences of the flux ratios between the components of
these two known lenses (Falco et al. 1999).}. We investigate whether
the large differences between the photometric flux ratios (see \S~4.1)
and the spectroscopic flux density ratio could be caused by the
continuum variations combined with the differential time delay. We
estimate the time delay between components A and B using the SIS model
(Peacock 1999)\footnote{We applied our SIS model to B 1600+434 whose
time delay is known to be about 50 days (Koopmans et al. 2000). Our
calculation gave 35 days for this lensed quasar --- in rough agreement with
the observation.}:
\begin{equation}
c{\Delta}t=(1+z_l){\frac{{D_s}{D_l}}{D_{ls}}}{{\alpha}_{E}}(\theta_{+}-\theta_{-}), 
\end{equation}
where $D_{s}$, $D_{l}$, $D_{ls}$, ${\alpha}_{E}$, and $z_{l}$ are the
angular size distances from observer to the source quasar, from
observer to the lensing galaxy, from the lensing galaxy to the source
quasar, the Einstein radius in arcseconds and the redshift of the
lensing galaxy, respectively. The observed separations from the center 
of the lensing galaxy are represented as ${\theta}_{+}$ and ${\theta}_{-}$. 
We suppose the redshift of the lensing galaxy is 0.4 (see below), the
velocity dispersion of the lensing galaxy is 230 km sec$^{-1}$ (see
below), $H_{0}=$70 km sec$^{-1}$ Mpc$^{-1}$, ${\Omega}_{m}=0.3$, and
${\Omega}_{\Lambda}=0.7$. Using these parameters, we determine a time
delay of about 15 days. Quasars generally do not experience large
variations, such as shown in Figure~4 (or Figure~5), in less than 15
days, therefore, there is a significant possibility that other
additional phenomena cause the differences between the two components
of SDSS~J0924+0219. 

One such effect might be microlensing. 
We cannot see clear evidence of microlensing in both spectra (Figure~4), 
but microlensing should be universal in quadruple lens (Witt, Mao, \& Schechter 1995),
therefore, it is natural that microlensing events happen to both components or to one of the
two components and cause the difference between the two components
(if we suppose the same situation of Q2237+0305 (Walsh, Carswell, \& Weymann 1979), 
the microlensing optical depth (Schmidt, Webster, \& Lewis 1998) supports this fact).
Furthermore, if there is another object which darkens component D or 
component D is a second lensing galaxy (see below discussion), it might have
an effect on the observed spectrum of component A, i.e. the reddening
of component A could be larger than that of component B. However, this 
extinction should be time invariant (effectively), and therefore, 
we should see the same extinction in the the photometric flux ratios, 
but we cannot see it. 
Confirming spectra and direct imaging are needed to determine whether 
the differences between the spectroscopic and direct imaging flux ratios are real
or an observational artifact.

Although the SIS model with an external shear (Kochanek 1991) predicts the existence
of a fourth lensed component whose amplification is as bright as the 
brightest of the other three lensed components, there are no stellar
components except components A, B, and C in all the images
(Figure~3). Using the positions of components A--C and 
of lensing galaxy G (which of course might not be the
only lensing galaxy) we can fit the SIS model with an external shear,
with a projected potential:
\begin{equation}
\psi(r,{\phi}) = {{\alpha}_{E}}r + {\gamma \over 2} r^2 \cos(2\phi - 2\phi_\gamma), 
\end{equation}
where ${{\alpha}_{E}}$ is the Einstein radius of the SIS model in
arcseconds, $r$ and $\phi$ are the radial and the angular parts,
respectively, of the angular position on the sky, and $\phi_\gamma$ is
the position angle of the shear, measured East of North. Fitting this
model to the positions (but not the fluxes) of components A--C, we get
${{\alpha}_{E}}=0\farcs850$, $\gamma=0.065$, and
$\phi_{\gamma}=173.65^\circ$, with a source position ($\Delta$R.A.,
$\Delta$Dec.) = ($0\farcs025$, $-0\farcs04$) relative to component
G. This value of ${{\alpha}_{E}}$ corresponds to a velocity dispersion
230 km sec$^{-1}$ with the estimated redshift of the lensing galaxy
(see final paragraph of this section). The observed
positions of components A--C, the observed position of component D and
the predicted positions of the lensed components are given in Table~5
below. With six constraints and five free parameters it is no
surprise that we obtain small residuals ($<0\farcs04$) from the
observed positions of components A--C. Although the flux ratios of
observed components were {\it not} taken as constraints, the predicted
flux ratios between components A and B and between components A and C
are in agreement with the observations. The predicted position of the
fourth lensed component (hereafter, component D$'$) is shown in
Figure~6a and Figure~6b. Figure~6a is the image plane of this model,
and Figure~6b is the enlarged part of the $i$ subtracted image of 
Figure~3 with the measured
positions of components A--C, component D and the predicted position
of component D$'$. Component D is separated by only about 0\farcs15
from the predicted position of component D$'$ (the filled circle on
Figure~6a or the open circle on Figure~6b). Also, the predicted
position of component D$'$ is within the region occupied by component
D (Figure~6b).

These results suggest that it is possible that either 1) component D
really is the ``missing'' fourth lensed component, or 2) that
component D is a object which is mixed component D$'$ (predicted fourth lensed component) 
with a foreground object 
that is obscuring the light from component D$'$. 
According to Schechter \& Wambsganss 2002, microlensing causes 
demagnificating (or vanishing) the fourth lensed image at ``saddlepoint''.
The difference between the expected 
magnitude and the observed magnitude is large 
($\Delta{m}=m_{expected}-m_{observed}=-2.7$), but this marginally consistent with
Figure~3 of Schechter \& Wambsganss 2002. Futhermore, some literatures reported 
that milli-lensing and/or micro-lensing produce anomalous flux ratios of 
four-image gravitationally lensed systems (Subramanian, Chitre, \& Narasimha 1985, 
Metcalf \& Madau 2001, Keeton 2003, and Kochanek \& Dalal 2003).
The second case is also likely. 
If there are a foreground dusty galaxy superposed upon component D$'$, 
it could be obscuring and reddening the light from component D$'$.
This might also explain why component D is
bluer than component G; component D may have some contribution from
component D$'$, which is expected to be relatively blue. The reduced 
${\chi}^{2}$ of component D from the PSF subtraction in the $i$ image is 20.91 
and it is comparable with the other stellar component (Table~1), therefore, 
there is a possibility that component D includes a stellar component. In addition to 
reddening by the foreground object, the demagnifying effect of microlensing might also 
darken component D$'$, and therefore, component D might be much fainter than 
component D$'$.

Two additional minor possibilities are that component D is a second lensing
galaxy and contributes significantly to the lensing potential, which
would change the lensing model such that the predicted position of
component D$'$ is incorrect, and that this lensing system has a very 
interesting lensing potential which produces only three lensed components. 
According to Keeton, Kochanek, \& Seljak (1997); Kassiola \& Kovner (1993);
and Wallington \& Narayan (1993), there are some cases that non-singular lensing 
potentials with large shears, large ellipticities, or large core radii produce only 
three lensed component. However, SDSS~J0924+0219 cannot be explained by these 
standard lensing models, such as non-singular lensing potentials with large shears because 
components of a ``standard'' three component lensing system are expected to be on the same 
side of the lensing galaxy in the non-singular lens models (Kassiola \& Kovner
1993; Wallington \& Narayan 1993) while the
three components of SDSS~J0924+0219 are not on the same side of the
presumed lensing galaxy (component G). 
Spectroscopy of image D is just one way for the question to be resolved.
Higher resolution imaging, say with HST, may resolve it without any need for
spectra.

According to Fukugita, Shimasaku, \& Ichikawa (1995), a typical elliptical galaxy at
$0.2<z<0.5$ has $1.3<{g-r}<1.8$ and $0.5<{r-i}<1.0$, which are close
to the estimated colors of component G (see Table~2). The estimated
colors and the spherical appearance of component G, therefore,
indicate that it may be an elliptical galaxy at $0.2<z<0.5$. This
would be the primary lensing galaxy of this lensing system. If the
redshift of the lensing galaxy is 0.4, the $1\farcs8$ separation of
components A and B requires the velocity dispersion of the lensing
galaxy to be about 230 km sec$^{-1}$ in the SIS model, while the
Faber-Jackson law predicts 225 km sec$^{-1}$ from the $i$ magnitude of
this galaxy ($i=$20.8). Here we assume that ${M_{i}}^{\ast}=-21.3$,
${\sigma}^{\ast}=225$ km sec$^{-1}$ (Blanton et al. 1999 and Kochanek
1996), ${\Omega}_{\rm M}=0.3$, $\Omega_{\Lambda}=0.7$, $H_{0}=70$ km
sec$^{-1}$ Mpc$^{-1}$, and the $K$-corrections of $i$ are 0.2 for a
$z=0.2$ elliptical galaxy and 0.4 for a $z=0.4$ (Inada 2001). This
result favors a redshift of 0.4 for the lensing galaxy, and a velocity
dispersion of about 230 km sec$^{-1}$.

\section{Summary and Conclusion}

Using a selection algorithm tuned to recover a previously discovered
lensed quasar (SDSS~1226$-$0006), we have identified several
additional lensed quasar candidates from the SDSS data. We confirmed
that one of them, SDSS~J0924+0219, is a lensed quasar with follow-up
observations using the Walter Baade 6.5-m and the Keck II
telescopes. The redshift of the source quasar is $z=1.524$. The
maximum separation is $1\farcs78$. The velocity difference between
component A and component B is very small, less than 100 km
sec$^{-1}$. We can directly see the lensing galaxy in the original
MagIC $i$ image, and we can see it more clearly in the PSF-subtracted
images. The estimated colors and the magnitudes of the lensing galaxy
are consistent with those of a typical elliptical galaxy at $z=0.4$
with a velocity dispersion of 230 km sec$^{-1}$. 
We can see only a faint red component near the
predicted position of the fourth lensed image (using the SIS model with an external shear). 
We consider that this faint red component is a fourth lensed component darkened and 
reddened by a foreground objects and microlensing.
To settle the issue of what causes the lack of the
fourth quasar component, we need to obtain deeper and higher
resolution images and fainter spectroscopy of component D. 

\acknowledgments 

Funding for the creation and distribution of the SDSS Archive has been
provided by the Alfred P. Sloan Foundation, the Participating
Institutions, the National Aeronautics and Space Administration, the
National Science Foundation, the U.S. Department of Energy, the
Japanese Monbukagakusho, and the Max Planck Society. The SDSS Web site
is http://www.sdss.org/. The SDSS is managed by the Astrophysical
Research Consortium (ARC) for the Participating Institutions. The
Participating Institutions are The University of Chicago, Fermilab,
the Institute for Advanced Study, the Japan Participation Group, The
Johns Hopkins University, Los Alamos National Laboratory, the
Max-Planck-Institute for Astronomy (MPIA), the Max-Planck-Institute
for Astrophysics (MPA), New Mexico State University, Princeton
University, the United States Naval Observatory, and the University of
Washington. G.~T.~R. and D.~P.~S. acknowledge support from National
Science Foundation grant AST99-00703 and D.~J.~E. acknowledges support from National
Science Foundation grant AST00-98577 and an Alfred P. Sloan Research Fellowship. 
Finally, we thank the staffs of
the Keck and Las Campanas Observatories for their excellent
assistance.

\clearpage

\begin{figure}
\plotone{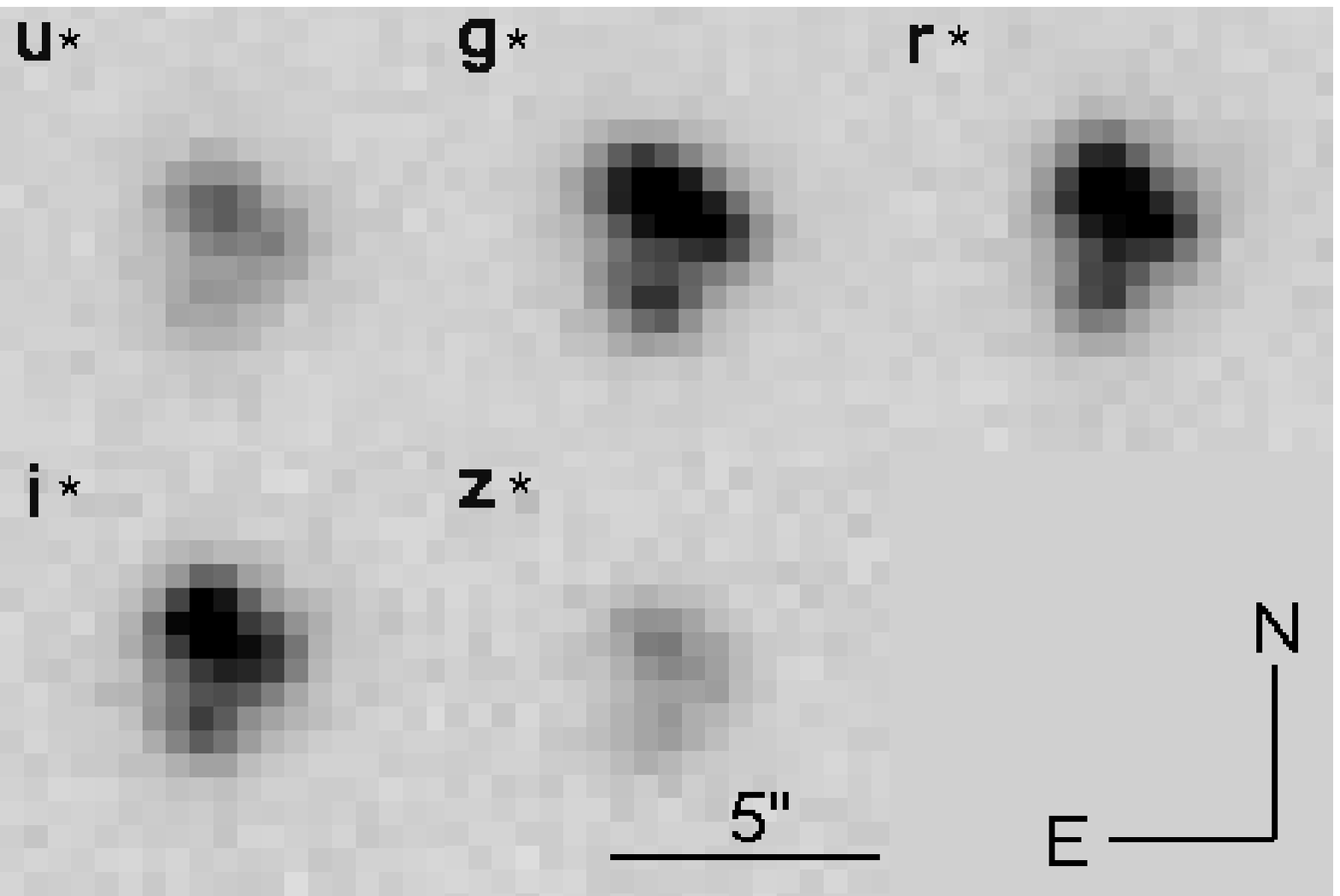}
\caption{The SDSS image of SDSS~J0924+0219 in all bands. The pixel size
is $0\farcs396$. The seeing of this field was $1\farcs4$. Total
magnitudes are 18.68, 18.43, 18.34, 18.09, and 17.98 in $u^*$, $g^*$,
$r^*$, $i^*$, and $z^*$, respectively.
\label{fig1}}
\end{figure}

\clearpage

\begin{figure}
\plotone{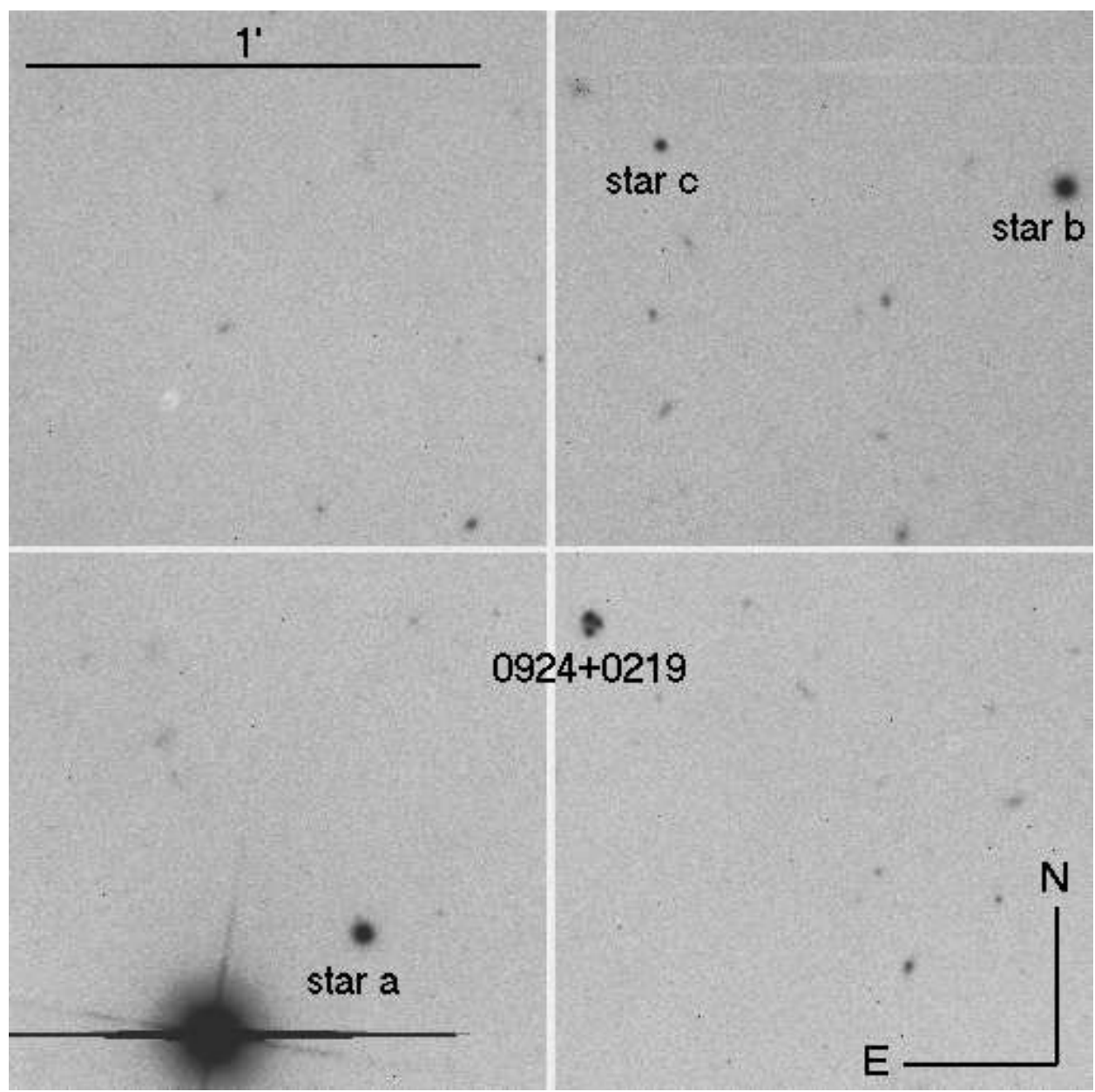}
\caption{The MagIC $r$ image of SDSS~J0924+0219 with nearby stars. This
CCD has four readout quadrants, which appear as four separate images
in this picture.
\label{fig2}}
\end{figure}

\clearpage

\begin{figure}
\plotone{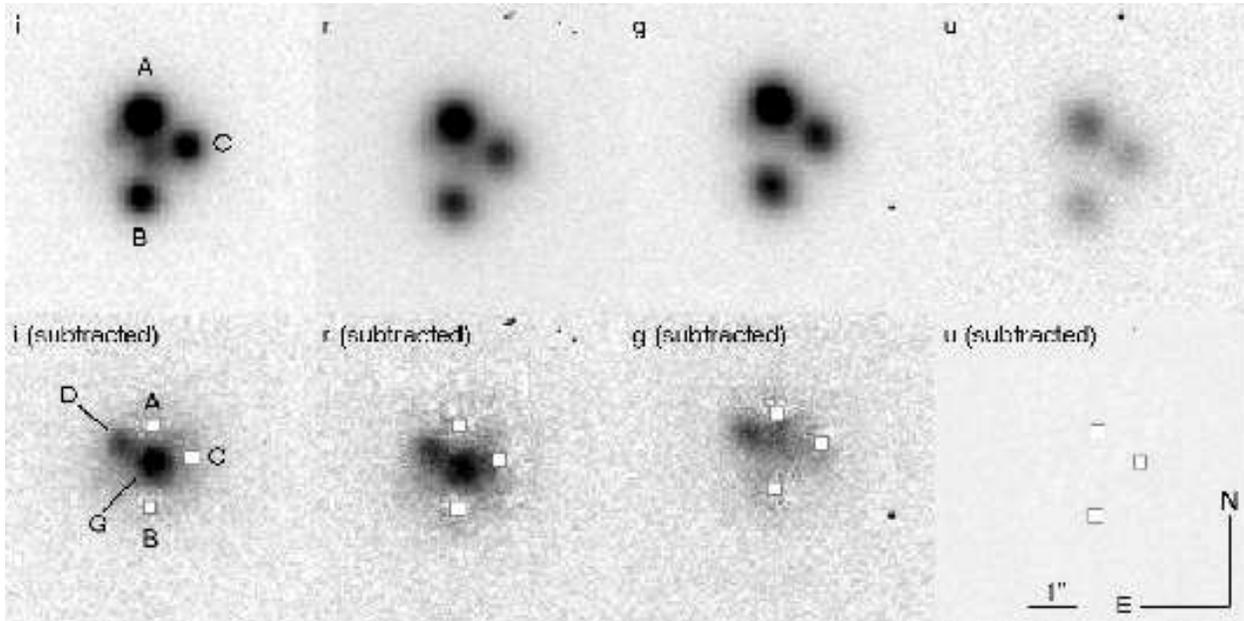}
\caption{The upper panels are the MagIC images of SDSS~J0924+0219 and
the lower panels are the PSF-subtracted images. The open squares in the
right panels represent the measured centers of components A--C. 
We can clearly see the three stellar components (A--C) in all original
images. We can also see the central extended object (G) and component D
at the upper left of component G in the $i$ original image (we can see them
more clearly in the PSF-subtracted image). We can see the central extended object 
(G) and the unknown fifth component (D) at the upper left of component G in the $i$, 
$r$ and $g$ PSF-subtracted image. However, we cannot see any other components in the 
PSF-subtracted image. The $i$, $r$ and $g$ magnitudes of component D are
estimated to be about 21.6, 21.8 and 22.5, respectively, and 
those of component G are estimated to be about 20.8, 21.3, and 22.7, respectively.
The pixel size of these images is $0\farcs069$, and the seeing was less than
$0\farcs75$. The exposure time was 300 seconds.
\label{fig3}}
\end{figure}

\clearpage

\begin{figure}
\plotone{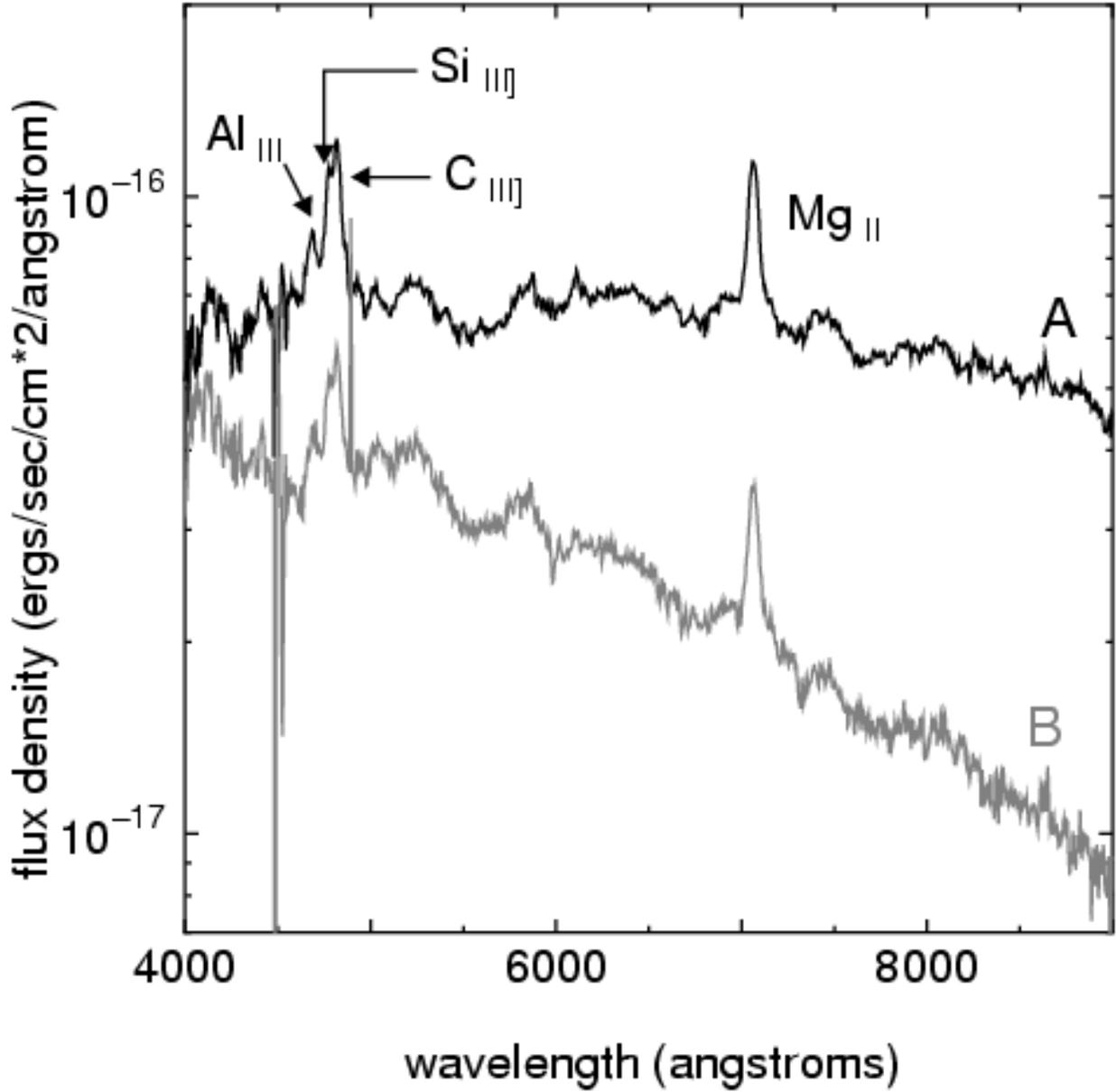}
\caption{Spectra of SDSS~J0924+0219 components A and B taken with ESI
on Keck II. Both components have \ion{Al}{3}, \ion{Si}{3]},
\ion{C}{3]}, and \ion{Mg}{2} emission line at $z=1.524$. There are
some bad columns around 4500{\,\AA} and 4900{\,\AA}.
\label{fig7}}
\end{figure}

\clearpage

\begin{figure}
\plotone{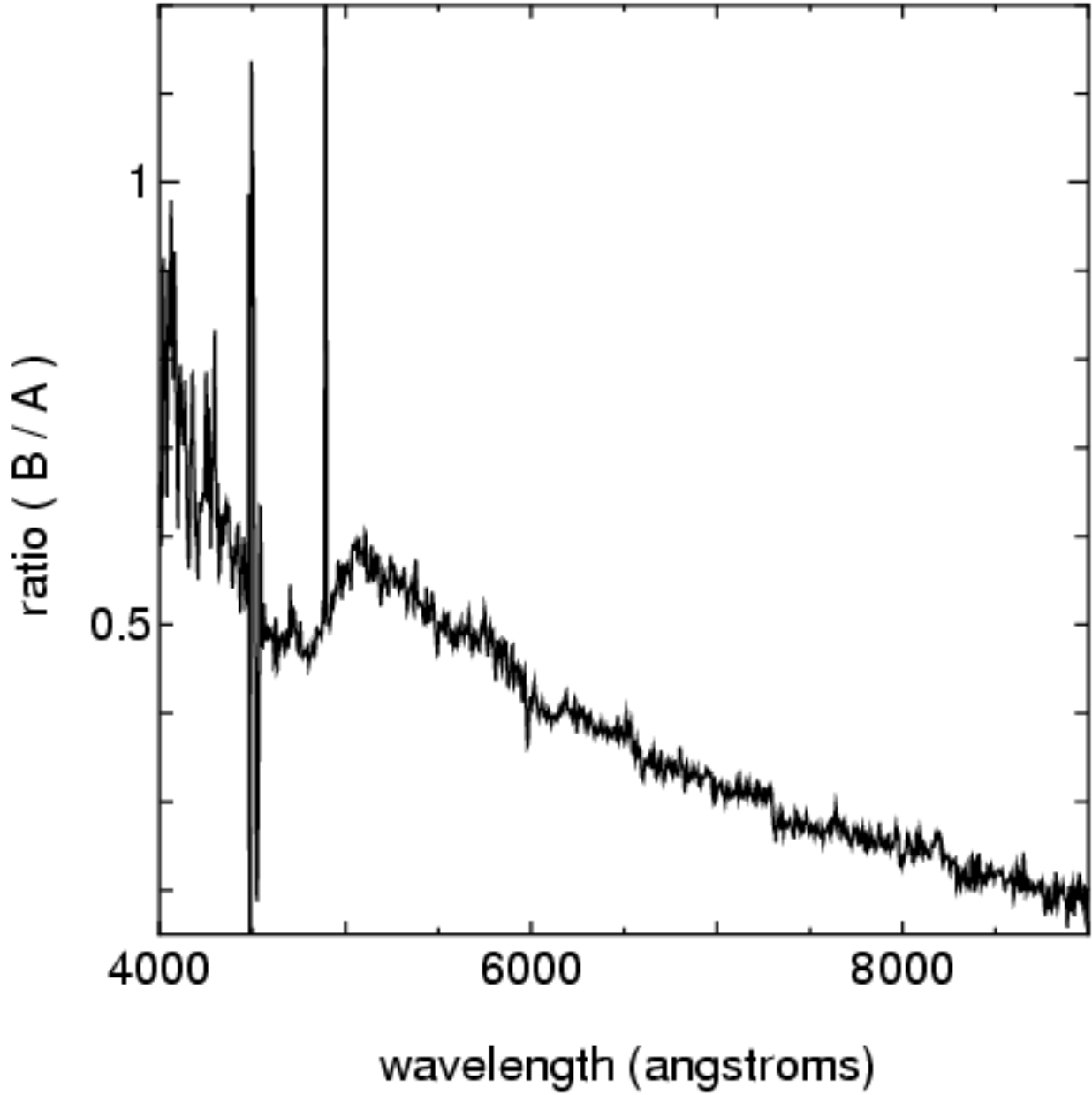}
\caption{The flux ratio between the two components (A and B) of
SDSS~J0924+0219. Although the flux ratio is not constant, it could be
explained by the continuum variations combined with the differential
time delay of each component and/or microlensing. There are some bad
columns around 4500{\,\AA} and 4900{\,\AA}.
\label{fig8}}
\end{figure}
\clearpage

\begin{figure}
\plottwo{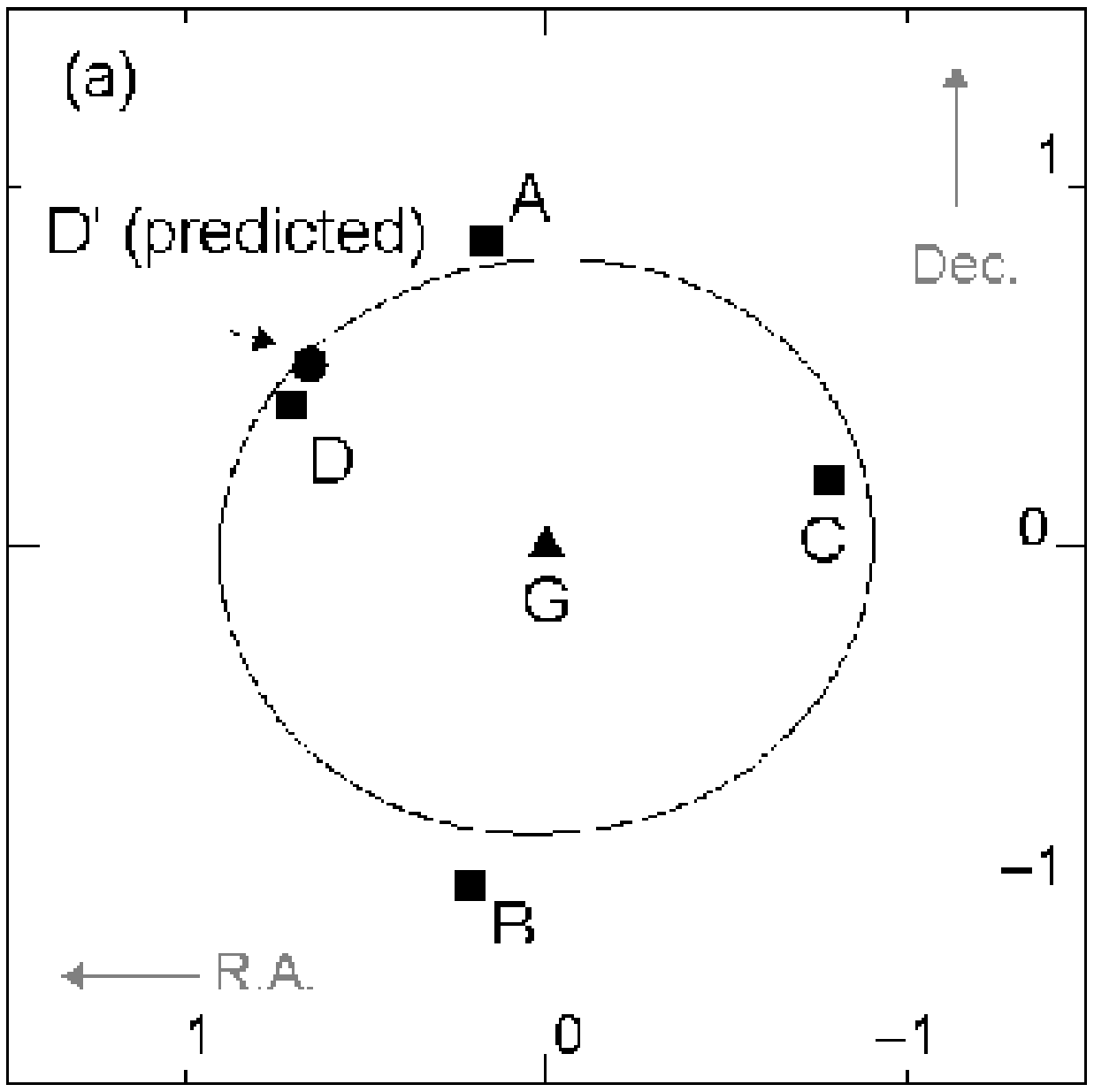}{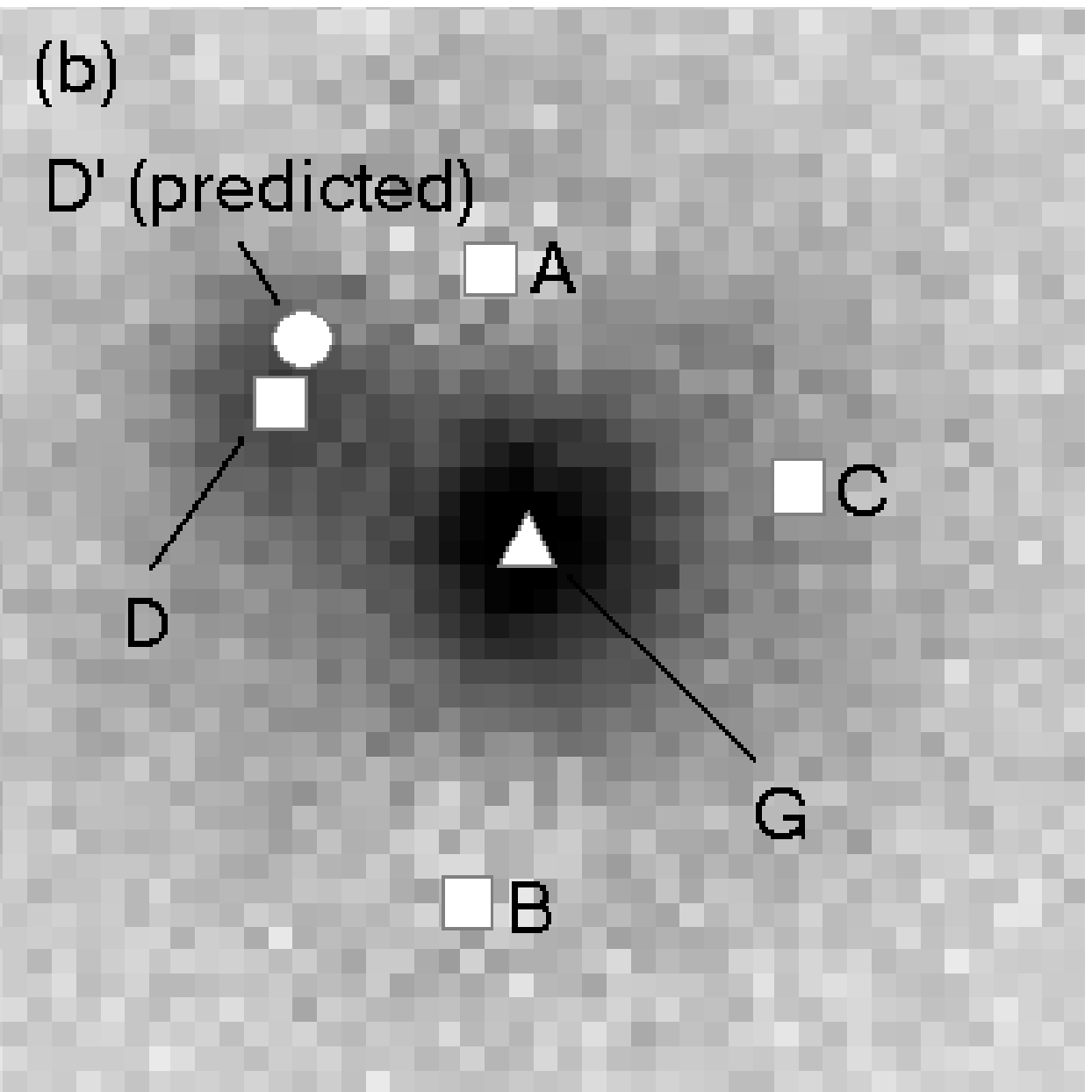}
\caption{ Positioning of the components relative to component G,
including the predicted location of a fourth lensed component, as
discussed in the text. ({\em Left}) The image plane of the model, the
filled circle represents the predicted position of component D$'$ and
the filled squares represent the measured position of components A--C
and component D. The filled triangle represents the measured position
of component G. ({\em Right}) An enlarged part of the $i$ subtracted image of Figure~3 (at the
same scale as the left panel). The white symbols in the right panel
correspond to the filled symbols in the left hand image. The
predicted position of component D$'$ was calculated using the SIS
model with an external shear restricted only by the positions of the
other three lensed components. The observed position of component D is
separated by only about $0\farcs15$ from the predicted position of
component D$'$, and the predicted position of component D$'$ is within
the contours of component D. Component D, however, is too red to be a
quasar (Table~2). Both panels are same size, $3\arcsec\times3\arcsec$
and are centered on the position of component G.
\label{fig9}}
\end{figure}

\clearpage

\begin{deluxetable}{crrrr} 
\tablecolumns{6} 
\tablewidth{0pc} 
\tablecaption{REDUCED ${\chi}^{2}$ FROM PSF SUBTRACTION} 
\tablehead{ 
\colhead{Object} & \colhead{$u$} & \colhead{$g$} & 
\colhead{$r$} & \colhead{$i$} }
\startdata 
Component A & 3.52 & 43.41 & 34.2 & 64.1 \\
Component B & 2.32 & 5.45 & 5.10 & 19.0 \\
Component C & 6.11 & 12.94 & 32.5 & 39.0 \\
\enddata
\end{deluxetable}

\clearpage

\begin{deluxetable}{cccccccc} 
\tablecolumns{6} 
\tablewidth{0pc} 
\tablecaption{PHOTOMETRY FOR SDSS~J0924+0219 AND NEARBY STARS} 
\tablehead{ 
\colhead{Object} & \colhead{$u$\tablenotemark{a}} & \colhead{$g$\tablenotemark{a}} & 
\colhead{$r$\tablenotemark{a}} & \colhead{$i$\tablenotemark{a}} & \colhead{$u-g$} & \colhead{$g-r$} & \colhead{$r-i$} }
\startdata 
Component A\tablenotemark{b} & 19.66$\pm$0.02 & 19.46$\pm$0.01 & 18.97$\pm$0.02 & 18.87$\pm$0.02 & 0.20 & 0.49 & 0.10 \\
Component B\tablenotemark{b} & 20.49$\pm$0.05 & 20.34$\pm$0.04 & 19.89$\pm$0.04 & 19.79$\pm$0.03 & 0.15 & 0.45 & 0.10 \\
Component C\tablenotemark{b} & 20.55$\pm$0.05 & 20.38$\pm$0.05 & 19.94$\pm$0.04 & 19.91$\pm$0.03 & 0.17 & 0.44 & 0.03 \\
Component D\tablenotemark{b,c} & $>$22.30\tablenotemark{d} & 22.45$\pm$0.12 & 21.82$\pm$0.06 & 21.61$\pm$0.05 & \nodata\phn & 0.63 & 0.21 \\
Component G\tablenotemark{b,c} & $>$22.30\tablenotemark{d} & 22.73$\pm$0.13 & 21.25$\pm$0.05 & 20.78$\pm$0.05 & \nodata\phn & 1.48 & 0.47 \\

\hline

Star a & 20.80$\pm$0.01 & 18.06$\pm$0.01 & 16.68$\pm$0.02 & 16.03$\pm$0.01 & 2.74 & 1.38 & 0.65 \\
Star b & 20.72$\pm$0.09 & 18.13$\pm$0.01 & 16.72$\pm$0.02 & 15.68$\pm$0.01 & 2.59 & 1.41 & 1.04 \\
Star c & 23.15$\pm$0.06 & 21.01$\pm$0.04 & 19.47$\pm$0.02 & 18.57$\pm$0.02 & 2.14 & 1.54 & 0.90 \\
\enddata
\tablenotetext{a}{We used stars a, b and c as photometric standard stars. 
The error bars of components A, B, C, D and G
do not include the photometric uncertainties of these standards. 
The magnitude of component D is estimated using the flux integrated in 
the 20 kpc (corresponding to 50 pixels (about $3\farcs7$) for $z=0.4$, supposing ${\Omega}=0.3$,
$\Omega_{\Lambda}=0.7$, and $H_{0}=70$ km sec$^{-1}$ Mpc$^{-1}$) diameter regions centered on each component location, 
after subtracting the other components.}
\tablenotetext{b}{A, B, C, G, and D are denoted in Figure~3.}
\tablenotetext{c}{Components G and D are not observed in the PSF-subtracted $u$ image.}
\tablenotetext{d}{The upper limits of $u$ of components D and G are estimated by integrating the flux 
in the regions expected by the PSF-subtracted $i$ image, after subtracting PSFs.}
\end{deluxetable}

\clearpage

\begin{deluxetable}{cccccccc} 
\tablecolumns{8} 
\tablewidth{0pc} 
\tablecaption{EMISSION LINES} 
\tablehead{ 
\colhead{} & \multicolumn{3}{c}{Component A} & \colhead{} & 
\multicolumn{3}{c}{Component B} \\ 
\cline{2-4} \cline{6-8} \\ 
\colhead{Element({\,\AA})} & \colhead{${\lambda}_{obs}$({\,\AA})} & \colhead{FWHM({\,\AA})} & \colhead{Redshift} & \colhead{} &
\colhead{${\lambda}_{obs}$({\,\AA})} & \colhead{FWHM({\,\AA})} & \colhead{Redshift} }
\startdata 
\ion{Al}{3}(1857.40) & 4684.75 & 38.9 & 1.5222$\pm$0.003 & & 4690.14 & 42.4 & 1.5251$\pm$0.003 \\
\ion{Si}{3]}(1892.03) & 4775.94 & 36.0 & 1.5242$\pm$0.002 & & 4779.39 & 43.6 & 1.5261$\pm$0.003 \\
\ion{C}{3]}(1908.73) & 4815.97 & 48.5 & 1.5231$\pm$0.001 & & 4818.40 & 45.4 & 1.5244$\pm$0.002 \\
\ion{Mg}{2}(2798.75) & 7063.51 & 61.2 & 1.5238$\pm$0.001 & & 7063.01 & 60.8 & 1.5236$\pm$0.001 \\
\enddata 
\end{deluxetable} 

\clearpage

\begin{deluxetable}{crrrr} 
\tablecolumns{6} 
\tablewidth{0pc} 
\tablecaption{ASTROMETRY FOR SDSS~J0924+0219 AND NEARBY STARS} 
\tablehead{ 
\colhead{Object} & \colhead{R.A.(J2000)\tablenotemark{a}} & \colhead{Dec.(J2000)\tablenotemark{a}} & 
\colhead{${\Delta}$R.A.(sec)\tablenotemark{b}} & \colhead{${\Delta}$Dec.($^{''}$)\tablenotemark{b}} }
\startdata 
Component A & 09 24 55.8293 & +02 19 25.356 & +0.0108 & +0.847 \\
Component B & 09 24 55.8327 & +02 19 23.565 & +0.0142 & $-$0.944 \\
Component C & 09 24 55.7659 & +02 19 24.691 & $-$0.0549 & +0.182 \\
Component D & 09 24 55.8653 & +02 19 24.897 & +0.0468 & +0.388 \\
Component G & 09 24 55.8185 & +02 19 24.509 & 0.0000 & 0.000 \\

\hline

Star a & 09 24 57.7258 & +02 18 44.566 & +1.9464 & $-$38.882 \\
Star b & 09 24 55.1770 & +02 20 25.660 & $-$0.6024 & +62.212 \\
Star c & 09 24 51.6797 & +02 20 19.861 & $-$3.9824 & +56.413 \\
\enddata
\tablenotetext{a}{
Astrometry is from Figure~3, based on the 
SDSS coordinates of stars a, b, and c.
The errors of components A, B, C, D, and G (not including the absolute errors 
of stars a, b, and c) are 0.018, 0.042, 0.046, 0.123, and 0.085 pixel per coordinate,
(0.0012, 0.0029, 0.0032, 0.0085, and 0.0059 arcsecond per coordinate,) respectively.}
\tablenotetext{b}{Position relative to component G}
\end{deluxetable}

\clearpage

\begin{deluxetable}{lccccccc} 
\tablecolumns{8} 
\tablewidth{0pc} 
\tablecaption{OBSERVED AND PREDICTED POSITIONS OF THE LENSED COMPONENTS} 
\tablehead{ 
\colhead{} & \multicolumn{3}{c}{observed} & \colhead{} & 
\multicolumn{3}{c}{predicted} \\ 
\cline{2-4} \cline{6-8} \\ 
\colhead{component} & \colhead{${\Delta}$R.A.($^{''}$)\tablenotemark{a}} & 
\colhead{${\Delta}$Dec.($^{''}$)\tablenotemark{a}} & 
\colhead{ratio(*/A)\tablenotemark{b}} & \colhead{} &
\colhead{${\Delta}$R.A.($^{''}$)\tablenotemark{a}} & 
\colhead{${\Delta}$Dec.($^{''}$)\tablenotemark{a}} & 
\colhead{ratio(*/A)\tablenotemark{b}} }
\startdata 
Component A & +0.162 & +0.847 & 1.00 & & +0.1688 & +0.8446 & 1.00 \\
Component B & +0.213 & $-$0.944 & 0.44\tablenotemark{d} & & +0.2209 & $-$0.9289 & 0.38 \\
Component C & $-$0.789 & +0.182 & 0.42\tablenotemark{d} & & $-$0.7548 & +0.1868 & 0.41 \\
Component D & +0.702 & +0.388 & 0.09\tablenotemark{d} & & \nodata\phn & \nodata\phn & \nodata\phn \\
Component D$'$ \tablenotemark{c} & \nodata\phn & \nodata\phn & \nodata\phn & & +0.6532 & +0.4962 & 0.85 \\
\enddata 
\tablenotetext{a}{Positions relative to component G}
\tablenotetext{b}{`` * '' represents each component, A, B, C, D, and D$'$}
\tablenotetext{c}{Predicted fourth lensed component}
\tablenotetext{d}{Mean ratio of all observed bands.}
\end{deluxetable}


\begin{thebibliography}{}
\bibitem[Blanton]{blanton01} Blanton. M. R., et al. 2001, \aj,
121, 2358
\bibitem[Chiba]{chiba99} Chiba, M. \& Yoshii, Y. 1999, \apj,
510, 42
\bibitem[Falco]{falco99} Falco, E. E., et al. 1999, \apj,
523, 617
\bibitem[Fukugita2]{fukugita96} Fukugita, M., Ichikawa, T., Gunn, J. E., Doi, M., Shimasaku, K., \& Schneider, D. P. 1996, \aj,
111, 1748
\bibitem[Fukugita3]{fukugita95} Fukugita, M., Shimasaku, K., \& Ichikawa, T. 1995, PASP,
107, 945
\bibitem[Fukugita4]{fukugita91} Fukugita, M. \& Turner, E. L. 1991, \mnras,
253, 99
\bibitem[Gunn]{gun98} Gunn, J. E., et al. 1998, \aj,
116, 3040
\bibitem[Hogg]{hogg01} Hogg, D. W., Finkbeiner, D. P., Schlegel,D. J., \& Gunn, J. E. 2001, \aj,
122, 2129
\bibitem[Inada1]{inada01} Inada, N. 2001, Master's thesis of University of Tokyo, Department of Physics.
\bibitem[Inada2]{inada02} Inada, N., et al. 2003, in preparation
\bibitem[Kassiola]{kassiola93} Kassiola, A. \& Kovner, I. 1993, \apj,
417, 450
\bibitem[Keeton]{keeton03} Keeton, C. R., 2003, \apj,
584, 664
\bibitem[Keeton]{keeton97} Keeton, C. R., Kochanek, C. S., \& Seljak, U. 1997, \apj,
482, 604
\bibitem[Kochanek]{kochanek03} Kochanek, C. S., \& Dalal, N. 2003, \apj, submitted (astro-ph/0302036)
\bibitem[Kochanek]{kochanek96} Kochanek, C. S., 1996, \apj,
466, 638
\bibitem[Kochanek]{kochanek91} Kochanek, C. S., 1991, \apj,
373, 354
\bibitem[Koopmans]{koopmans00} Koopmans, L. V. E., de Bruyn, A. G., Xanthopoulos, E., \& Fassnacht, C. D. 2000, A{\&}A,
356, 391
\bibitem[Lupton]{lupton01} Lupton, R. H., Gunn, J. E., Ivezi\'{c}, \'{Z}., Knapp, G. R., Kent, S., \& Yasuda, N. 2001, ADASS X,
2001adass-10, 269
\bibitem[Metcalf]{metcalf01} Metcalf, B. R., Madau, P, 2001, \apj,
563, 9
\bibitem[Peacock]{peacock99} Peacock, J. A., 1999, Cambridge University Press,
``Cosmological Physics''
\bibitem[Pier]{pier02} Pier, J. R., Munn, J. A., Hindsley, R. B., Hennessy, G. S., Kent, S. M., Lupton, R. H., \& Ivezi\'{c}, \'{Z}. 2002, \apj, submitted
\bibitem[Refsdal]{refsdal64} Refsdal, S. 1964, \mnras,
128, 307 
\bibitem[Richards]{richards02} Richards, G. T., et al. 2002, \aj,
123, 2945
\bibitem[Richards]{richards01} Richards, G. T., et al. 2001, \aj,
122, 1151
\bibitem[Schechter]{schechter98} Schechter, P. L., Gregg, M. D., Becker, R. H., Helfand, D. J., \& White, R. L. 1998, \aj,
115, 1371 
\bibitem[Schechter]{schechter02} Schechter, L. P. \& Wambsganss. 2002, \apj, submitted (astro-ph/0204425)
\bibitem[Schneider]{schneider02} Schneider, D. P., et al. 2002, \aj,
123, 567 
\bibitem[Smith]{smith02} Smith, A., et al. 2002, \aj,
123, 2121
\bibitem[Stoughton]{stoughton02} Stoughton, C., et al. 2002, \aj,
123, 485
\bibitem[Subramanian]{subramanian85} Subramanian, K., Chitre, S. M., \& Narasimha, D. 1985, \apj,
289, 37
\bibitem[Sutin]{sutin97} Sutin, B. M. 1997, Proc. SIPE,
2871, 1116
\bibitem[Wallington]{wallington93} Wallington, S. \& Narayan, R. 1993, \apj,
403, 517
\bibitem[Trevese]{trevese01} Tr\'{e}vese, D., Kron, R. G., \& Bunone, A. 2001, \apj,
551, 103
\bibitem[Turner]{turner84} Turner, E. L., Ostriker, J. P., \& Gott, J. R., III 1984, \apj,
284, 1
\bibitem[Wisotzki]{wisotzki95} Wisotzki, L., Koehler, T., Ikonomou, M., \& Reimers, D. 1995, A{\&}A,
297, L59 
\bibitem[Witt]{witt95} Witt, H. J., Mao, S., \& Schechter, L. P. 1995, \apj,
443, 18
\bibitem[York]{york00} York, D. G., et al. 2000, \aj,
120, 1579
\end{thebibliography}
\end{document}